# Achievable Degrees of Freedom for Closed-form Solution to Interference Alignment and Cancellation in Gaussian Interference Multiple Access Channel

Qu Xin[†] and Chung G. Kang[††]

[†] UNISOC, Shanghai, China; email: xin.qu@unisoc.com
[††] School of Electrical Engineering, Korea University, Seoul, Korea; email: ccgkang@korea.ac.kr

*Abstract*—A combined technique of interference alignment (IA) and interference cancellation (IC), known as interference alignment and cancellation (IAC) scheme, has been proposed to improve the total achievable degrees of freedom (DoFs) over IA. Since it is NP-hard to solve the transceiver under a given tuple of DoFs or to maximize the total achievable DoFs in the general system configuration by IA (or IAC), the optimal transceiver cannot be obtained in polynomial time. Meanwhile, it has been known that a closed-form yet suboptimal transceiver can be designed for IAC by employing a symbol-to-symbol (STS) alignment structure. As its performance has not been known yet, we aim to derive the total DoFs that can be achieved by such suboptimal but closed-form IAC transceivers for Gaussian interference multiple access channels with $K$ receivers and $J$ users (transmitters), each with $M$ antennas. Our analysis shows that the closed-form IAC transceivers under consideration can achieve a maximum total achievable DoFs of $2M$, which turns out to be larger than those achieved in classical IA, e.g., $2MK/(K+1)$ DoFs by a specific configuration where each link has the same target DoFs. Moreover, considering the NP-hardness of deriving the maximum total achievable DoFs with the optimal IAC transceiver, its upper bound has been derived for comparison with the results of our closed-form IAC transceiver. Numerical results illustrate that its performance can be guaranteed within 20% of the upper bound when the number of multiple access channels are relatively small, e.g., $K \leq 4$.

## I. Introduction

THE total degrees of freedom (DoFs) achieved by vector space interference alignment (IA) depends on two aspects: the maximum number of interference signals caused by single interfering source and the aligned level for interference signals from all interfering sources. With the limited number of Tx/Rx antennas, in order to improve the achievable total DoFs over IA, a vector space interference alignment and cancellation (IAC) technique has been proposed, which combines vector space IA with interference cancellation (IC). As IC operation can immediately eliminate interferences at a receive side, it makes more signal subspace than IA under the same number of Tx/Rx antennas. Therefore, IAC is expected to gain more DoFs over IA.

With limited spatial dimension, IAC is not always feasible. To determine the total achievable DoFs, one needs to address the feasibility of IAC first. IAC is feasible when the interferences are properly reduced so that enough interference-free space can be saved for the desired signals. Both IA and IC operations contribute to eliminating interferences. With IC operation at receive side, the decoded signal packets at one receiver are sent to other receivers over backhaul link, and therefore, the interferences caused by such known signals can be immediately extracted and do not burden on the interference subspace. Consequently, the feasibility of IAC only depends on the feasibility of IA under the IAC-specific interference scenario.

There are several works that have studied the feasibility of IA for $K$-user interference channel with $M$ transmit antennas and $N$ receiver antennas [1-3]. In general, they related the feasibility issue to the problem of determining the solvability of a set of quadratic multivariate polynomial equations which can be summarized as follows:

$$\mathbf{U}_k^H \mathbf{H}_{kj} \mathbf{V}_j = \mathbf{0}, \quad \forall k \neq j \qquad (1)$$

$$\text{rank}\left(\mathbf{U}_k^H \mathbf{H}_{kk} \mathbf{V}_k\right) = d_k \qquad (2)$$

where $\mathbf{V}_j$ denotes a precoding matrix at transmitter $j$, $\mathbf{U}_k$ denotes a zero-forcing matrix at receiver $k$, and $\mathbf{H}_{kj}$ denotes a channel matrix from transmitter $j$ to receiver $k$, and $d_k$ denotes the target DoFs to obtain at receiver $k$. The transceivers $\{\mathbf{V}_j\}$ and $\{\mathbf{U}_k\}$, $j,k \in [1,K]$, are the unknowns to be solved from (1) and (2).

In the existing works [2, 3], (1) and (2) can be solved only for symmetric system configurations where each Tx/Rx pair has the same target DoF $d_k = d$ and the same number of Tx/Rx antennas, i.e., $M/N$, which is divisible by $d$. For general system configurations with a given tuple of DoFs $(d_1, d_2, \cdots, d_K)$, it has been shown that solving (1) and (2) is NP-hard, as long as each node is equipped with at least 3 antennas [4]. Therefore, a few heuristic algorithms have been proposed to solve IA transceivers in an iterative manner [5][6]. However, these algorithms cannot determine whether the solutions exist, nor is there any guarantee for converging to the optimal solutions even when they exist. Subsequently, the total

DoFs achieved by these heuristic algorithms is not guaranteed. Furthermore, it has been also shown that finding the maximum total DoFs achieved by IA is also NP-hard for the general system configuration [4].

Considering that it is impossible to derive the optimal transceivers or the maximum total achievable DoFs until now, we aim to derive some suboptimal yet definite results which do not depend on the performance of iterative algorithms. In our earlier work [7], a symbol-to-symbol alignment (STS) structure has been proposed to IAC in a Gaussian interference MAC channel. Based on the proposed STS structure, a closed-form IAC solution can be obtained, allowing us to determine the total achievable DoFs definitely. Furthermore, the necessary and sufficient conditions for the existence of closed-form IAC solutions have also been derived. In this paper, we aim to investigate the total DoFs that can be achieved by such closed-form solutions. Since it is NP-hard to find the maximum total achievable DoFs for optimal IAC solutions, our derived result has been compared with the upper bound on achievable DoFs of IAC, which can be obtained from the infeasibility IAC conditions.

Section II presents the system model and the concepts of IAC. Section III reviews the proposed IAC with STS alignment scheme as well as its closed-form solutions in [7]. Section IV derives the total DoFs for the Gaussian interference MAC system achieved by the IAC with STS alignment scheme. Furthermore, the related analytical results, including the upper bound on achievable DoF of IAC, are presented along with the solid proofs. Finally, Section V provides the concluding remarks.

## II. PRELIMINARIES

### A. System Model

Fig. 1 illustrates a general configuration of a $(K,M,J)$ Gaussian interference MAC system. There are $K$ interfering MACs, in each of which a receiver of the $k$-th MAC system is associated with a group of $N_k$ users ( $N_k \geq 1$ and $\sum_{k=1}^{K} N_k = J \geq K$ ), assuming that each receiver and user are equipped with $M$ antennas. For notational convenience, we denote the $j$-th user in the $k$-th MAC by user[j,k]. Each user[j,k] desires to send $d^{[j,k]}$ independent concurrent signal packets to its target receiver $k$ ($d^{[j,k]} \leq M$). We can further define the total DoFs in the system as $DoF = \sum_{k=1}^{K}\sum_{j=1}^{N_k} d^{[j,k]}$.

Let $\mathbf{x}^{[j,k]}$ denote the transmitted signal vector of dimension $d^{[j,k]} \times 1$ from the user[j,k] ( $j=1,2,\cdots,N_k$; $k=1,2,\cdots,K$ ), in which each element of the vector corresponds to one independent signal packet, denoted by $\{x_\ell^{[j,k]}\}_{\ell=1}^{d^{[j,k]}}$. Furthermore, the $M \times M$ channel matrix from user[j',k'] to receiver $k$ is given by $\mathbf{H}_k^{[j',k']}$, with each entry drawn independently from a continuous distribution, while allowing no channel extension. Let $\mathbf{V}^{[j,k]} = \begin{bmatrix} \mathbf{v}_1^{[j,k]} & \mathbf{v}_2^{[j,k]} & \cdots & \mathbf{v}_{d^{[j,k]}}^{[j,k]} \end{bmatrix}$ represent an $M \times d^{[j,k]}$ transmit precoding matrix at user[j,k], where each column vector is applied to each signal packet, and $\mathbf{U}_k$ denote an $M \times \left(\sum_{j=1}^{N_k} d^{[j,k]}\right)$ zero-forcing matrix at receiver $k$. Subsequently, the output signal vector of dimension $\left(\sum_{j=1}^{N_k} d^{[j,k]}\right) \times 1$ at receiver $k$ can be represented as

$$\mathbf{y}_k = \underbrace{\mathbf{U}_k^H \sum_{j=1}^{N_k} \mathbf{H}_k^{[j,k]} \sum_{\ell=1}^{d^{[j,k]}} \mathbf{v}_\ell^{[j,k]} x_\ell^{[j,k]}}_{desired\ signals}$$
$$+ \underbrace{\mathbf{U}_k^H \sum_{k'=1, k'\neq k}^{K} \sum_{j'=1}^{N_{k'}} \mathbf{H}_k^{[j',k']} \sum_{\ell'=1}^{d^{[j',k']}} \mathbf{v}_{\ell'}^{[j',k']} x_{\ell'}^{[j',k']}}_{int\ erferences} + \mathbf{U}_k^H \mathbf{n}_k$$

(3)

where $\mathbf{n}_k$ is a zero-mean additive white Gaussian noise vector, $\mathbf{n}_k \sim \mathcal{CN}(\mathbf{0}, \sigma^2 \mathbf{I})$. Throughout this paper, $\mathbf{A}^H$ denotes the conjugate transpose operator for a matrix $\mathbf{A}$.

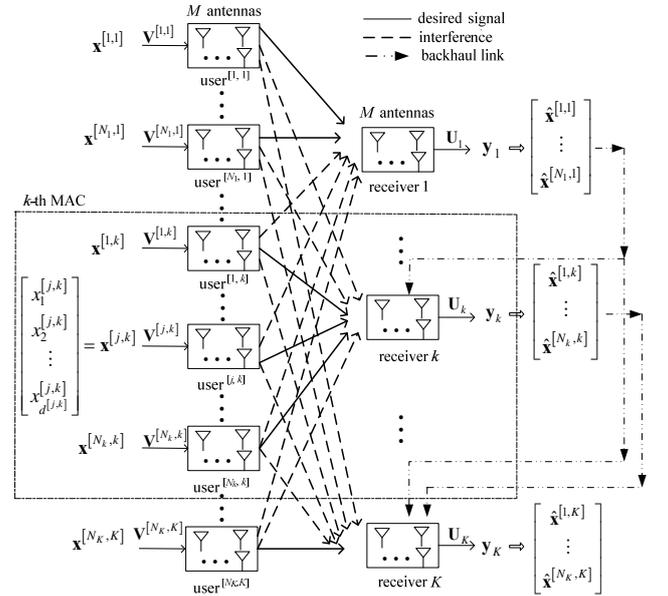

Fig. 1. $(K,M,J)$ Gaussian interference MAC system model

### B. Interference alignment and cancellation (IAC)

In the sequel, we briefly introduce how IAC works in the MAC channel. At the transmit side, IA operation works to precode the signals jointly, so that the interferences caused at the receive side can be effectively aligned. At the receive side, IC operation performs successively, i.e., the decoded signals

$\left\{ \hat{\mathbf{x}}^{[j,k]} \middle| j \in [1, N_k] \right\}$ at receiver $k$ are sent over a backhaul link to other receivers which have not been decoded yet for cancellation. Therefore, a decoding operation is supposed to be performed in one receiver at a time. Without loss of generality, we assume that the decoding order follows from receiver 1 to $K$. When the receiver $k$ decodes $\mathbf{y}_k$, it has already obtained $\left\{ \hat{\mathbf{x}}^{[j',k']} \middle| j' \in [1, N_{k'}], k' \in [1, k-1] \right\}$ from earlier decoding, so that the corresponding interferences $\sum_{k'=1}^{k-1} \mathbf{H}_k^{[j',k']} \mathbf{V}^{[j',k']} \hat{\mathbf{x}}^{[j',k']}$ can be immediately subtracted from $\mathbf{y}_k$. As a result, such part of interferences is not necessarily to be handled by IA operation. In other words, at each $k$-th MAC, IA operation takes charge of only the part of interferences caused by the $(k+1)$-th to $K$-th MACs.

Additionally, note that the total number of received packets decreases as successive cancellation is performed. In particular, there exist a total of $\sum_{k'=1}^{K} \sum_{j=1}^{N_{k'}} d^{[j',k']}$ packets initially at the receiver $k$, which has only a total of $\sum_{k'=k}^{K} \sum_{j=1}^{N_{k'}} d^{[j',k']}$ remaining packets by the time it gets its turn to decode $\mathbf{y}_k$. If $\sum_{k'=k}^{K} \sum_{j=1}^{N_{k'}} d^{[j',k']} \leq M$, the desired signals and interferences are separable with $M$ antennas, indicating that the interferences incurred by the $k$-th MAC are not required to be aligned. This further works at the $(k+1)$-st to $K$-th MACs, since the total number of remaining packets at each of them is no more than the one at the $k$-th MAC. In order to identify the receivers in which the interferences are not necessarily to be aligned, we allocate a new label $k_{\text{IAC}}$ to $(k-1)$-st MAC if the $k$-th MAC is the first one which holds $\sum_{k'=k}^{K} \sum_{j=1}^{N_{k'}} d^{[j',k']} \leq M$. In other words, IA operation is applied only to align the interferences incurred at the 1st to the $k_{\text{IAC}}$-th MACs. Furthermore, $\sum_{k'=k_{\text{IAC}}+1}^{K} \sum_{j=1}^{N_{k'}} d^{[j',k']} \leq M$ also indicates that the interferences caused by $\left\{ \mathbf{x}^{[j',k']} \middle| j' \in [1, N_{k'}], k' \in [k_{\text{IAC}}+1, K-1] \right\}$ do not require to be cancelled. Therefore, only the decoded signals at the 1st to the $k_{\text{IAC}}$-th MACs are required to be sent through the backhaul link for a cancellation purpose.

According to the system model, the signal subspace $\tilde{S}_k$ at $k$-th MAC is spanned by a set of signal vectors, $S_k = \left\{ \mathbf{H}_k^{[j,k]} \mathbf{v}_\ell^{[j,k]} \middle| \ell \in [1, d^{[j,k]}], j \in [1, N_k] \right\}$, while the interference subspace $\tilde{I}_k$ is spanned by a set of interference vectors, $I_k = \left\{ \mathbf{H}_k^{[j',k']} \mathbf{v}_{\ell'}^{[j',k']} \middle| \ell' \in [1, d^{[j',k']}], \forall j' \in [1, N_{k'}], k' \in [1, K], k' \neq k \right\}$

Moreover, $I_k$ can be also written in the union of two sets, i.e., $I_k = I_k^{\text{IA}} \cup I_k^{\text{IC}}$, where
$I_k^{\text{IA}} = \left\{ \mathbf{H}_k^{[j',k']} \mathbf{v}_{\ell'}^{[j',k']} \middle| \ell' \in [1, d^{[j',k']}], \forall j' \in [1, N_{k'}], k' \in [k+1, K] \right\}$
denotes the set of interferences aligned by IA operation and
$I_k^{\text{IC}} = \left\{ \mathbf{H}_k^{[j',k']} \mathbf{v}_{\ell'}^{[j',k']} \middle| \ell' \in [1, d^{[j',k']}], \forall j' \in [1, N_{k'}], k' \in [1, k-1] \right\}$
denotes the set of interferences that are eliminated by IC operation. Inspired by (1) and (2), the zero-forcing constraints on IAC can be given as

$$\mathbf{U}_k^H \mathbf{H}_k^{[j',k']} \mathbf{V}^{[j',k']} = \mathbf{0} \quad (4)$$

$$\text{rank}\left(\mathbf{U}_k^H \mathbf{S}_k\right) = \sum_{j=1}^{N_k} d^{[j,k]} \quad (5)$$

for $\forall k \in [1, k_{\text{IAC}}]$, $\forall k' \in [k+1, K]$, and $\forall j' \in [1, N_{k'}]$, where $\mathbf{S}_k = \left[ \mathbf{H}_k^{[1,k]} \mathbf{V}^{[1,k]} \cdots \mathbf{H}_k^{[N_k,k]} \mathbf{V}^{[N_k,k]} \right]$. (4) guarantees that at each $k$-th MAC, all interferences $\mathbf{H}_k^{[j',k']} \mathbf{V}^{[j',k']} \in I_k^{\text{IA}}$ are aligned onto the subspace that is orthogonal to $\mathbf{U}_k$, while (5) ensures that the signal subspace at the $k$-th MAC has dimension $\sum_{j=1}^{N_k} d^{[j,k]}$ and is linearly independent of the interference subspace.

## III. INTERFERENCE ALIGNMENT AND CANCELLATION: CLOSED-FORM SOLUTIONS

In this section, we review the proposed IAC with an STS alignment scheme, which was originally introduced in [7] as a specific alignment scheme. The main works have been summarized, including the proposed IAC graph, which allows for obtaining the closed-form solutions, along with the necessary and sufficient conditions for the existence of closed-form IAC solutions.

### A. IAC with STS alignment structure

Since solving the transceivers, $\left\{ \mathbf{V}^{[j,k]} \right\}$ and $\left\{ \mathbf{U}_k \right\}$ jointly from the set of quadratic equations in (4) is NP-hard, we consider designing the transmitters $\left\{ \mathbf{V}^{[j,k]} \right\}$ first and then the receivers $\left\{ \mathbf{U}_k \right\}$ in our earlier work. In order to satisfy both (4) and (5), $\left\{ \mathbf{V}^{[j,k]} \right\}$ can be found by aligning interferences in $I_k$, such that the interference subspace $\tilde{I}_k$ is complimentary to the signal subspace $\tilde{S}_k$. And then, the receivers $\left\{ \mathbf{U}_k \right\}$ can always be found in the left null space of $\tilde{I}_k$. As the interferences in $I_k^{\text{IC}}$ can be cancelled and thus, they do not burden the dimension of $\tilde{I}_k$, only the interferences in $I_k^{\text{IA}}$ are required to be aligned.

The above discussion motivates us to seek for an approach that can align the interferences in $I_k^{\text{IA}}$, such that $\tilde{I}_k$ is linearly independent of $\tilde{S}_k$ while satisfying the following condition:

$$\dim(\tilde{S}_k) + \dim(\tilde{I}_k) \leq M \quad (6)$$

where $\dim(\mathcal{A})$ corresponds to the cardinality of a basis for a vector space $\mathcal{A}$.

Suppose $\bar{I}_k^{\text{SIA}} = \{\bar{i}_{1,k}, \cdots, \bar{i}_{n,k}, \cdots, \bar{i}_{Z_k,k}\} \in \mathbb{C}^{M \times Z_k}$ denote a set of basis vectors that span $\tilde{I}_k$ with $|\bar{I}_k^{\text{SIA}}| = Z_k = M - \dim(\tilde{S}_k)$. Therefore, $\{\mathbf{H}_k^{[j'',k'']} \mathbf{v}_{\ell''}^{[j'',k'']} \mid \forall \mathbf{H}_k^{[j'',k'']} \mathbf{v}_{\ell''}^{[j'',k'']} \in I_k^{\text{IA}}\}$ should be aligned onto $\bar{I}_k^{\text{SIA}}$, resulting in the following set of alignment equations at receiver $k$:

$$\mathbf{H}_k^{[j'',k'']} \mathbf{v}_{\ell''}^{[j'',k'']} = \alpha_{(\ell''-1) \cdot Z_k + 1, k}^{[j'',k'']} \bar{i}_{1,k} + \cdots \\ + \alpha_{(\ell''-1) \cdot Z_k + n, k}^{[j'',k'']} \bar{i}_{n,k} + \cdots + \alpha_{\ell'' \cdot Z_k, k}^{[j'',k'']} \bar{i}_{Z_k,k} \quad (7)$$

where $\{\alpha_{(\ell''-1) \cdot Z_k + n, k}^{[j'',k'']}\}$ are the coefficients that represent the relative magnitude of vectors.

Concerning the form of $\{\alpha_{(\ell''-1) \cdot Z_k + n, k}^{[j'',k'']}\}$ and $\{\bar{i}_{n,k}\}$ in (7), there are two cases to be addressed: 1) both $\{\alpha_{(\ell''-1) \cdot Z_k + n, k}^{[j'',k'']}\}$ and $\{\bar{i}_{n,k}\}$ are unknowns; 2) $\{\alpha_{(\ell''-1) \cdot Z_k + n, k}^{[j'',k'']}\}$ are assumed as the arbitrary coefficients and $\{\bar{i}_{n,k}\}$ are unknowns. In [7], however, we have shown that both cases cannot guarantee the independency of $\tilde{I}_k$ and $\tilde{S}_k$. Furthermore, we have proposed the STS alignment structure in (7), which ensures the independency of $\tilde{I}_k$ and $\tilde{S}_k$ while satisfying (6). The STS structure includes two parts: first, $\bar{I}_k^{\text{SIA}}$ is constructed by selecting the interference vectors in $I_k^{\text{IA}}$, i.e., $\bar{I}_k^{\text{SIA}} \subset I_k^{\text{IA}}$, and then, $I_k^{\text{IA}} - \bar{I}_k^{\text{SIA}} = \{\hat{i}_{1,k}, \cdots, \hat{i}_{f,k}, \cdots, \hat{i}_{|I_k^{\text{IA}}| - Z_k, k}\} \subset I_k^{\text{IA}}$; second, each $\hat{i}_{f,k} \in I_k^{\text{IA}} - \bar{I}_k^{\text{SIA}}$ is uniquely aligned on to one $\bar{i}_{n,k}$, giving only one coefficient $\alpha_{(\ell''-1) \cdot Z_k + n, k}^{[j'',k'']}$ to be non-zero in (7), i.e., $\mathbf{H}_k^{[j'',k'']} \mathbf{v}_{\ell''}^{[j'',k'']} = \alpha_{(\ell''-1) \cdot Z_k + n, k}^{[j'',k'']} \bar{i}_{n,k}$. Replacing $\bar{i}_{n,k}$ with the detailed expression, a set of linear alignment equations that are formed at receiver $k$ can be equivalently rewritten as

$$\Phi_k \triangleq \left\{ \text{span}\left(\mathbf{H}_k^{[j',k']} \mathbf{v}_{\ell'}^{[j',k']}\right) = \text{span}\left(\mathbf{H}_k^{[j'',k'']} \mathbf{v}_{\ell''}^{[j'',k'']}\right) \mid \\ \mathbf{H}_k^{[j',k']} \mathbf{v}_{\ell'}^{[j',k']} \in \bar{I}_k^{\text{SIA}}, \mathbf{H}_k^{[j'',k'']} \mathbf{v}_{\ell''}^{[j'',k'']} \in I_k^{\text{IA}} - \bar{I}_k^{\text{SIA}} \right\} \quad (8)$$

where $k \in [1, K]$ and $k'' \neq k'$ assures that the interferences from the same transmitter are not aligned. Then, a complete set of linear alignment equations for system is given by $\Psi \triangleq \{\Phi_k\}_{k=1}^K$.

### B. Closed-form solutions

The closed-form transmitters $\{\mathbf{V}^{[j,k]}\}$ can be solved from $\Psi$. In [7], an IAC graph $\mathcal{G} = (\mathcal{P}, \mathcal{E})$ has been proposed to analyze the solvability of $\Psi$, where $\mathcal{P}$ and $\mathcal{E}$ represent a set of vertices and a set of edges, respectively. Furthermore, we have shown that $\mathcal{G}$ and $\Psi$ have one-to-one correspondence, i.e., each precoding vector $\mathbf{v}_\ell^{[j,k]}$ and each alignment equation in $\Psi$ can be represented by one unique vertex and one unique edge in $\mathcal{G}$, respectively, and vice versa. Note that IAC graph is different from a general representation of graph, as each edge in IAC graph holds a label to declare the index of MAC where the alignment equation is formed. Moreover, as each variable $\mathbf{v}_\ell^{[j,k]}$ may not appear in all alignment equations in $\Psi$, the equations involving the same subset of variables has been collected into one subset. In other words, $\Psi$ can be divided into several independent subsets, and solving the independent subsets respectively is equivalent to solving $\Psi$. Correspondingly, each independent subset of $\Psi$ forms one independent connected subgraph of $\mathcal{G}$. Each connected subgraph has no isolated vertex, and no connection to other subgraphs. For each connected subgraph $q$, it is denoted as $\mathcal{G}_q = (\mathcal{P}_q, \mathcal{E}_q)$, such that $\mathcal{G} = \{\mathcal{G}_1, \mathcal{G}_2, \cdots, \mathcal{G}_Q\}$, $|\mathcal{P}| = \sum_{q=1}^Q |\mathcal{P}_q|$, and $|\mathcal{E}| = \sum_{q=1}^Q |\mathcal{E}_q|$, where $\mathcal{P}_q$ and $\mathcal{E}_q$ represent a set of vertices and a set of edges in subgraph $q$, respectively. By analyzing IAC graph $\mathcal{G}$, the necessary and sufficient conditions for solving $\Psi$ has been derived in [7], and can be summarized in the below:

**Proposition 1.** For a connected subgraph in IAC graph $\mathcal{G}$, if and only if the vertices form at most one loop, the precoding vectors involved can always be solved.

Then, it has been further proven that the solutions $\{\mathbf{v}_\ell^{[j,k]}\}$, obtained from $\Psi$, can always guarantee the independence of $\tilde{S}_k$ and $\tilde{I}_k$, which can be stated in the following proposition:

**Proposition 2.** The precoding vectors $\{\mathbf{v}_\ell^{[j,k]}\}$ solved from Proposition 1 can always lead to the independence of $\tilde{S}_k$ and $\tilde{I}_k$, $k \in [1, K]$.

Finally, a necessary and sufficient condition for the existence of closed-form IAC transceivers has been proven, and can be summarized by the following theorem:

**Theorem 1.** In the $(K, M, J)$ Gaussian interference MAC

system, based on the proposed symbol-to-symbol alignment scheme, closed-form IAC solutions exist if and only if the following inequalities are satisfied:

$$\sum_{j=1}^{N_k} d^{[j,k]} + \max\left\{d^{[j',k']}\right\}_{\substack{j'\in[1,N_{k'}]\\k'\in[k+1,K]}} \leq M \quad (9)$$

$$\sum_{k'=k_{\text{IAC}}+1}^{K}\sum_{j'=1}^{N_{k'}} d^{[j',k']} \leq M, \quad k_{\text{IAC}}=1,2,\cdots,K-1 \quad (10)$$

$$\sum_{j=1}^{N_1} d^{[j,1]} + \sum_{k=1}^{k_{\text{IAC}}}(k-1)\sum_{j=1}^{N_k} d^{[j,k]} + \sum_{k'=k_{\text{IAC}}+1}^{K}(k_{\text{IAC}}-1)\sum_{j'=1}^{N_{k'}} d^{[j',k']} \leq k_{\text{IAC}} \cdot M \quad (11)$$

At last, as long as the conditions in **Theorem 1** are satisfied, the closed-form IAC solutions exist and can be found from IAC graph $\mathcal{G}$. Since the interferences caused by the 1st MAC can be cancelled immediately at the 2nd to the $K$-th MACs, $\left\{\mathbf{v}_\ell^{[j,1]}\middle|\forall\ell\in\left[1,d^{[j,k]}\right],\forall j\in[1,N_k]\right\}$ will not appear in $\Psi$. Let $\mathbf{V} = \left\{\mathbf{v}_\ell^{[j,k]}\middle|\forall\ell\in\left[1,d^{[j,k]}\right],\forall j\in[1,N_k],\forall k\in[2,K]\right\}$ denote the set of precoding vectors in $\Psi$ as well as the set of vertices in $\mathcal{G}$. Therefore, each subset $\mathbf{V}_q \in \mathbf{V}$ can be found by going through "loop" in its corresponding subgraph $\mathcal{G}_q$, $q\in[1,Q]$ and $\mathbf{V}=(\mathbf{V}_1,\mathbf{V}_2,\cdots,\mathbf{V}_Q)$. For the given system configuration $(K,M,J)$ and $\left\{d^{[j,k]}\middle|\forall j\in[1,N_k],\forall k\in[1,K]\right\}$, construction of the basis $\bar{I}_k^{\text{SIA}}$ as well as selection of the alignment pairs for each equation in (8) is not unique. Therefore, there may exist many different possible $\Psi$, each of which leads to a different $\mathcal{G}$. As the expression of the closed-form solutions depend on the structure of $\mathcal{G}$, it is difficult to explicitly give a general expression which can cover all the possibilities. Hence, we briefly give the procedures to obtain the closed-form solutions for each possibility in the below. More details on the explicit expressions of closed-form solutions for the different possibilities have been given in [8].

According to **Proposition 1**, there are two types of subgraphs: one-loop case and no-loop case. For one-loop case, one precoding vector involved in the loop can be expressed by a product of itself and one full rank matrix, and thus, this precoding vector can be chosen as any eigenvector of the full rank matrix. Then, the remaining precoding vectors represented in $\mathcal{G}_q$ can be computed through the edge. For the no-loop case, we can just pick the direction of one precoding vector randomly and find the remaining precoding vectors through the edges. Furthermore, $\left\{\mathbf{v}_\ell^{[j,1]}\middle|\forall\ell\in\left[1,d^{[j,k]}\right],\forall j\in[1,N_k]\right\}$ can be found by letting $\mathbf{V}^{[j,1]} \subset \left(\mathbf{H}_1^{[j,1]}\right)^{-1}\left(\tilde{I}_1\right)^{\perp}$ where $\tilde{I}_1$ is known from

$\mathbf{V}$. Once the transmitters $\left\{\mathbf{v}_\ell^{[j,k]}\middle|\forall\ell\in\left[1,d^{[j,k]}\right],\forall j\in[1,N_k],\forall k\in[1,K]\right\}$ are solved, $\{\tilde{I}_k\}_{k=1}^{K}$ is determined and then, the receiver $\mathbf{U}_k$ can be obtained by $\mathbf{U}_k \subset \left(\tilde{I}_k\right)^{\perp}$, $\forall k\in[1,K]$.

## IV. THE ACHIEVABLE DoFs WITH THE CLOSED-FORM IAC SOLUTIONS

In this section, we first derive the DoFs that can be achieved for $(K,M,J)$ MAC system with the proposed suboptimal yet closed-form IAC solutions, denoted as $DoF_{\text{IAC}}^{\text{CS}}$. As it is NP-hard to find the maximum achievable DoFs by IAC, we then derive a necessary condition that must be satisfied by any tuple of DoF, $\left\{d^{[j,k]}\middle|\forall j\in[1,N_k],\forall k\in[1,K]\right\}$ for IAC, based on which an upper bound on the maximum achievable DoFs, denoted as $DoF_{\text{IAC}}^{\text{upper}}$, can be obtained.

### A. The achievable DoFs: $DoF_{\text{IAC}}^{\text{CS}}$

**Theorem 2.** In the $(K,M,J)$ Gaussian interference MAC system, the maximum achievable DoFs with the symbol-to-symbol alignment-based closed-form IAC solutions is $2M$ and it is achieved when $k_{\text{IAC}} = 2$.

*Proof:* The achievable DoFs can be derived from the necessary and sufficient conditions in Theorem 1. Substituting $DoF_{\text{IAC}}^{\text{CS}} = \sum_{k=1}^{K}\sum_{j=1}^{N_k} d^{[j,k]}$ into (11) yields

$$\sum_{k=1}^{k_{\text{IAC}}}(k-1)\sum_{j=1}^{N_k} d^{[j,k]} + (k_{\text{IAC}}-1)\left(DoF_{\text{IAC}}^{\text{CS}} - \sum_{k=1}^{k_{\text{IAC}}}\sum_{j=1}^{N_k} d^{[j,k]}\right) \leq k_{\text{IAC}} \cdot M - \sum_{j=1}^{N_1} d^{[j,1]} \quad (12)$$

After some algebraic manipulations, (12) can be equivalently expressed as

$$(k_{\text{IAC}}-1)\cdot\left(DoF_{\text{IAC}}^{\text{CS}} - M\right) - \sum_{k=2}^{k_{\text{IAC}}}\left(\sum_{k'=1}^{k}\sum_{j'=1}^{N_{k'}} d^{[j',k']} - \sum_{j=1}^{N_k} d^{[j,k]}\right) \leq M - \sum_{j=1}^{N_1} d^{[j,1]} \quad (13)$$

We can then rewrite (13) as

$$\sum_{k=2}^{k_{\text{IAC}}}\left\{\left(DoF_{\text{IAC}}^{\text{CS}} - \sum_{k'=1}^{k}\sum_{j'=1}^{N_{k'}} d^{[j',k']}\right) - \left(M - \sum_{j=1}^{N_k} d^{[j,k]}\right)\right\} \leq M - \sum_{j=1}^{N_1} d^{[j,1]} \quad (14)$$

Referring to the STS alignment structure, the number of linear alignment equations formed at receiver $k$ is

$$|\Phi_k| = |\mathcal{I}_k^{\text{IA}}| - |\bar{\mathcal{I}}_k^{\text{SIA}}| = \left(DoF_{\text{IAC}}^{\text{CS}} - \sum_{k'=1}^{k}\sum_{j'=1}^{N_{k'}} d^{[j',k']}\right) - \left(M - \sum_{j=1}^{N_k} d^{[j,k]}\right)$$

and hence, (14) becomes

$$\sum_{k=2}^{k_{\text{IAC}}} |\Phi_k| \leq M - \sum_{j=1}^{N_1} d^{[j,1]} \quad (15)$$

Partitioning $\sum_{k=2}^{k_{\text{IAC}}} |\Phi_k|$ as $\sum_{k=2}^{k_{\text{IAC}}} |\Phi_k| = |\Phi_2| + \sum_{k=3}^{k_{\text{IAC}}} |\Phi_k|$ and substituting $|\Phi_2| = DoF_{\text{IAC}}^{\text{CS}} - M - \sum_{j=1}^{N_1} d^{[j,1]}$ into (15) yields

$$DoF_{\text{IAC}}^{\text{CS}} \leq 2M - \sum_{k=3}^{k_{\text{IAC}}} |\Phi_k| \quad (16)$$

For the last term on the right-hand side, as $\sum_{k=3}^{k_{\text{IAC}}} |\Phi_k| \geq 0$, $DoF_{\text{IAC}}^{\text{CS}}$ is bounded by $2M$ and furthermore, the bound is tight when $\sum_{k=3}^{k_{\text{IAC}}} |\Phi_k| = 0$. In fact, $2M$ can be achieved when $\sum_{k=3}^{k_{\text{IAC}}} |\Phi_k| = 0$, together with $|\Phi_2| \neq 0$, implying that $k_{\text{IAC}} = 2$. This completes the proof. ∎

In the sequel, using IAC graph $\mathcal{G} = (\mathcal{P}, \mathcal{E})$, we provide an intuitive interpretation on the proof. For a general $(K, M, J)$ MAC system, if we increase the total number of transmitted data streams $DoF_{\text{IAC}}^{\text{CS}}$, the caused total number of interferences will grow much faster than $DoF_{\text{IAC}}^{\text{CS}}$. Correspondingly, the number of edges grows much faster than that of the vertices in IAC graph $\mathcal{G}$. According to **Proposition 1**, in order for the existence of closed-form solutions $\{\mathbf{V}^{[j,k]} | \forall j \in [1, N_k], \forall k \in [1, K]\}$, $\mathcal{G}$ is allowed to have at most one loop at each subgraph $\mathcal{G}_q$, i.e., it requires that $|\mathcal{E}_q| \leq |\mathcal{P}_q|$, $q \in [1, Q]$. Therefore, $DoF_{\text{IAC}}^{\text{CS}}$ cannot be increased arbitrarily and $2M$ can be achieved when $|\mathcal{E}_q| = |\mathcal{P}_q|$ holds, $q \in [1, Q]$.

Assume that $\mathcal{G}$ is structured by following the same order as the decoding operation in IAC, namely, from the 1st to the $K$-th MACs. At the 1st MAC, each interference vector $\bar{t}_{f,1} \in \mathcal{I}_1^{\text{IA}} - \bar{\mathcal{I}}_1^{\text{SIA}}$ will be uniquely aligned onto one interference basis vector $\bar{t}_{n,1} \in \bar{\mathcal{I}}_1^{\text{SIA}}$ and moreover, the alignment operation is not allowed between any two basis vectors. Therefore, a set of vertices which correspond to the interference vectors aligned onto $\bar{t}_{n,1}$ have been connected to the reference vertex represented by $\bar{t}_{n,1}$, forming one connected subgraph with no loop. Then, a total of $|\bar{\mathcal{I}}_1^{\text{SIA}}| = M - \sum_{j=1}^{N_1} d^{[j,1]}$ subgraphs have been formed, each of which has $|\mathcal{E}_n| = |\mathcal{P}_n| - 1$, $n \in [1, (M - \sum_{j=1}^{N_1} d^{[j,1]})]$. Hence, each subgraph allows only one more edge and then, the total number of edges added through the 2nd to $K$-th MACs should be no more than $(M - \sum_{j=1}^{N_1} d^{[j,1]})$, giving the inequality in (15).

Next, we briefly illustrate how $2M$ is achieved when $k_{\text{IAC}} = 2$. Recall that the total number of edges added through the 2nd to $K$-th MACs should be no more than $(M - \sum_{j=1}^{N_1} d^{[j,1]})$. Together with $\sum_{k=3}^{k_{\text{IAC}}} |\Phi_k| = 0$ for $k_{\text{IAC}} = 2$, then, the following inequality should be satisfied:

$$\sum_{k=2}^{K} |\Phi_k| = |\Phi_2| \leq M - \sum_{j=1}^{N_1} d^{[j,1]} \quad (17)$$

Substituting $k_{\text{IAC}} = 2$ into (10) yields $\sum_{k=3}^{K} \sum_{j=1}^{N_k} d^{[j,k]} \leq M$. Let $\sum_{k=3}^{K} \sum_{j=1}^{N_k} d^{[j,k]} = M$ and then, we have

$$\begin{aligned} |\Phi_2| &= |\mathcal{I}_2^{\text{IA}}| - |\bar{\mathcal{I}}_2^{\text{SIA}}| \\ &= \left(\sum_{k=3}^{K}\sum_{j=1}^{N_k} d^{[j,k]}\right) - \left(M - \sum_{j=1}^{N_2} d^{[j,2]}\right) \\ &= \sum_{j=1}^{N_2} d^{[j,2]} \end{aligned} \quad (18)$$

Substituting (18) into (17), we have $\sum_{j=1}^{N_1} d^{[j,1]} + \sum_{j=1}^{N_2} d^{[j,2]} \leq M$. We can set $\sum_{j=1}^{N_1} d^{[j,1]} + \sum_{j=1}^{N_2} d^{[j,2]} = M$, together with $\sum_{k=3}^{K} \sum_{j=1}^{N_k} d^{[j,k]} = M$. Then, a total of $DoF_{\text{IAC}}^{\text{CS}} = 2M$ can be achieved.

*B. Upper bound on the maximum achievable DoFs with IAC: $DoF_{\text{IAC}}^{\text{upper}}$*

IAC inherits the NP-hardness from IA on maximizing the total achievable DoFs for the general system configurations. In this subsection, we first give a necessary condition for IAC feasibility in MAC channel, which is an extension of the existing results in [2][3] and [9]. Based on the necessary feasibility condition of IAC, we then derive an upper bound on the total DoFs that can be achieved with IAC, denoted as $DoF_{\text{IAC}}^{\text{upper}}$.

**Theorem 3.** *In the $(K, M, J)$ Gaussian interference MAC system, any tuple of DoFs $\{d^{[j,k]} | \forall j \in [1, N_k], \forall k \in [1, K]\}$ that is achievable with IAC must satisfy the following inequalities:*

$$\sum_{j=1}^{N_k} d^{[j,k]} + d^{[j',k']} \leq M, \quad \forall k, k', k' \neq k, \forall j' \in [1, N_{k'}] \quad (19)$$

$$\sum_{k'=k_{\text{IAC}}+1}^{K} \sum_{j'=1}^{N_{k'}} d^{[j',k']} \leq M, \quad k_{\text{IAC}} \leq K-1 \quad (20)$$

$$\sum_{k:\{\cdot,[j',k']\}\in\Upsilon_{\text{sub}}} \left(\sum_{j=1}^{N_k} d^{[j,k]}\right)\left(M - \sum_{j=1}^{N_k} d^{[j,k]}\right)$$
$$+ \sum_{[j',k']:\{k,\cdot\}\in\Upsilon_{\text{sub}}} d^{[j',k']}\left(M - d^{[j',k']}\right) \quad (21)$$
$$\geq \sum_{\{k,[j',k']\}\in\Upsilon_{\text{sub}}} \left(\sum_{j=1}^{N_k} d^{[j,k]}\right) d^{[j',k']}$$

where $\Upsilon_{\text{sub}}$ is a subset of indices defined by $\Upsilon = \{\{k,[j',k']\}: k\in[1,k_{\text{IAC}}], k'\in[k+1,K], j'\in[1,N_{k'}]\}$, and $\{\cdot,[j',k']\}$ or $\{k,\cdot\}$ denote that there exists $k$ or $[j',k']$ such that $\{k,[j',k']\} \in \Upsilon_{\text{sub}}$.

*Proof*: We first prove the inequality in (19). Referring to the MAC channel given in system descriptions, we have $\text{rank}(\mathbf{U}_k) = \sum_{j=1}^{N_k} d^{[j,k]}$ and $\text{rank}(\mathbf{V}^{[j',k']}) = d^{[j',k']}$. As each element of channel matrix $\mathbf{H}_k^{[j',k']}$ drawn i.i.d. from a continuous distribution, $\mathbf{H}_k^{[j',k']}$ is generic and hence, $\text{rank}(\mathbf{H}_k^{[j',k']}\mathbf{V}^{[j',k']}) = d^{[j',k']}$. Together with $\mathbf{U}_k \mathbf{H}_k^{[j',k']}\mathbf{V}^{[j',k']} = \mathbf{0}$, the transmitter $\mathbf{U}_k$ must satisfy $\text{span}(\mathbf{U}_k) \perp \text{span}(\mathbf{H}_k^{[j',k']}\mathbf{V}^{[j',k']})$ and thus, (19) is true. The inequality (20) is obvious by the definition of $k_{\text{IAC}}$, i.e., for any feasible IAC strategy, there always exists a $k_{\text{IAC}}$. Now we prove (21) for IAC in MAC channel by extending the existed proof for IA in [2][3] and [9], the total number of scalar variables should be no less than the total number of scalar equations for any subset $\Upsilon_{\text{sub}}$. For any certain $\{k,[j',k']\} \in \Upsilon_{\text{sub}}$, there is $\mathbf{U}_k^H \mathbf{H}_k^{[j',k']}\mathbf{V}^{[j',k']} = \mathbf{0}$. Recall $\text{rank}(\mathbf{U}_k) = \sum_{j=1}^{N_k} d^{[j,k]}$ and $\text{rank}(\mathbf{V}^{[j',k']}) = d^{[j',k']}$. Then, the total numbers of scalar variables in $\mathbf{U}_k$ and $\mathbf{V}^{[j',k']}$, after eliminating superfluous variables, are $\left(\sum_{j=1}^{N_k} d^{[j,k]}\right)\left(M - \sum_{j=1}^{N_k} d^{[j,k]}\right)$ and $d^{[j',k']}(M - d^{[j',k']})$, respectively. The details of the superfluous variables can be referred to [1] and [3]. As any combination of one column vector in $\mathbf{U}_k$ and one column vector in $\mathbf{V}^{[j',k']}$ forms one equation, the total number of scalar equations is $\left(\sum_{j=1}^{N_k} d^{[j,k]}\right) d^{[j',k']}$. As the interferences received at the $(k_{\text{IAC}}+1)$-th to $K$-th MACs are not required to be aligned in IAC, $\Upsilon$ only counts the 1st to $k_{\text{IAC}}$-th MACs, i.e.,

$\forall k: \{\cdot,[j',k']\} \in [1, k_{\text{IAC}}]$. Furthermore, since the interferences caused by 1st to $k_{\text{IAC}}$-th MACs can be immediately cancelled at $k$-th MAC, $\Upsilon$ only includes the remaining interferences, i.e., $\forall [j',k']: \{k,\cdot\}$, where $k' \in [k+1,K], j' \in [1,N_{k'}]$. This completes the proof of (21). ∎

In (21), there exist a total of $\left(2^{\sum_{k=1}^{k_{\text{IAC}}} \sum_{k'=k+1}^{K} N_{k'}}\right) - 1$ subsets, each of which has to be checked to identify the achievability of a tuple of DoFs $\{d^{[j,k]} | \forall j \in [1,N_k], \forall k \in [1,K]\}$. Then, the underlying computational complexity will be prohibitive and thus, the upper bound given by Theorem 3 is hard to obtain. Meanwhile, if we let $\Upsilon_{\text{sub}} = \Upsilon$, the following infeasibility condition can be configured from (21): any tuple of DoFs $\{d^{[j,k]} | \forall j \in [1,N_k], \forall k \in [1,K]\}$ is not achievable if

$$\sum_{k=1}^{k_{\text{IAC}}} \left(\sum_{j=1}^{N_k} d^{[j,k]}\right)\left(M - \sum_{j=1}^{N_k} d^{[j,k]}\right) + \sum_{k=2}^{K} \sum_{j=1}^{N_k} d^{[j,k]}\left(M - d^{[j,k]}\right)$$
$$< \sum_{k=1}^{k_{\text{IAC}}} \left\{\left(\sum_{j=1}^{N_k} d^{[j,k]}\right)\left(\sum_{k'=k+1}^{K} \sum_{j'=1}^{N_{k'}} d^{[j',k']}\right)\right\} \quad (22)$$

Then, an upper bound on the maximum achievable DoFs can be derived by solving (22) with equality. However, it is not straightforward to obtain the analytical expression of $DoF_{\text{IAC}}^{\text{upper}}$, and so a numerical solution of $DoF_{\text{IAC}}^{\text{upper}}$ is given for comparison in next subsection.

*C. Comparison of DoFs: $DoF_{\text{IAC}}^{\text{CS}}$ vs. $DoF_{\text{IAC}}^{\text{upper}}$*

In this subsection, the maximum achievable DoFs of the proposed closed-form IAC solutions, given by $DoF_{\text{IAC}}^{\text{CS}} = 2M$, is compared with the upper bound of the general IAC, denoted as $DoF_{\text{IAC}}^{\text{upper}}$. Numerical results are given to show the comparison between $DoF_{\text{IAC}}^{\text{CS}}$ and $DoF_{\text{IAC}}^{\text{upper}}$. We first estimate $DoF_{\text{IAC}}^{\text{upper}}$ from (22). In $(K,M,J)$ MAC channel, we fix $K$ and $M$ while generating $\{N_k | k \in [1,K]\}$ and $\{d^{[j,k]} | j \in [1,N_k], k \in [1,K]\}$ randomly by following the discrete uniform distributions, $N_k \sim \mathcal{U}(1,M)$ and $d^{[j,k]} \sim \mathcal{U}(1,\text{int}(M/N_k))$, respectively. Notice that $J$ can be obtained by $J = \sum_{k=1}^{K} N_k$. For the tuple of DoFs $\{d^{[j,k]} | \forall j \in [1,N_k], \forall k \in [1,K]\}$, if (22) does not hold true, we have $DoF_{\text{IAC}}^{\text{upper}} = \sum_{k=1}^{K} \sum_{j=1}^{N_k} d^{[j,k]}$.

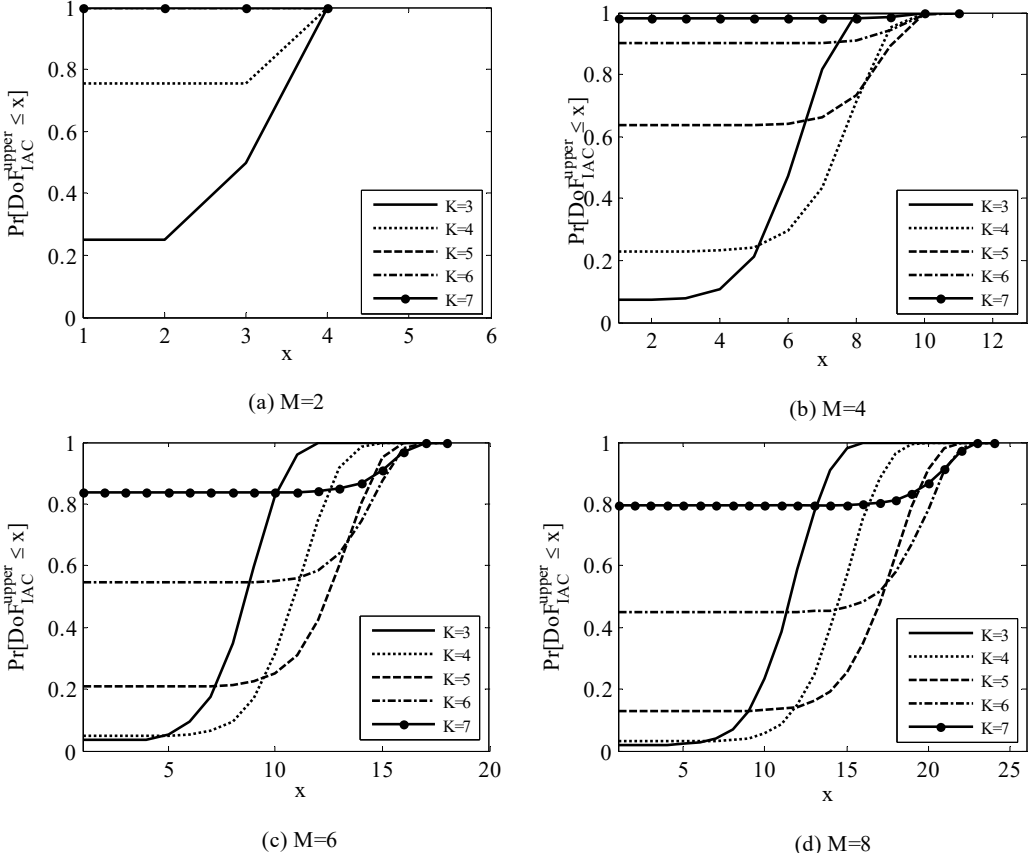

Fig 2. The CDF of $DoF_{IAC}^{CS}$ as varying $M$ and $K$

We calculate the cumulative distribution function (CDF) of $DoF_{IAC}^{upper}$ by $10^4$ Monte Carlo runs. In each run, $k_{IAC}$ can be calculated by its definition. As the interferences require to be aligned only when no fewer than three cells interfere with each other, we set $K \geq 3$. Furthermore, we consider $K \leq 7$ by assuming that each cell cooperates with the cells only in its first tier in cellular network.

Fig. 2 shows the CDFs of $DoF_{IAC}^{upper}$, $F(x) = \Pr(DoF_{IAC}^{upper} \leq x)$, for $M = 2, 4, 6, 8$. Note that there is no CDF curves for $K = 5, 6, 7$ in Fig. 2 (a), because with only two Tx/Rx antennas, IAC cannot be achieved by more than 4 MACs. Since $F(x) = \Pr(DoF_{IAC}^{upper} \leq x)$, the upper bound of DoFs for general IAC is given as $(DoF_{IAC}^{upper})^* = F^{-1}(1)$. It is observed from Fig. 2 that the upper bound is given by $(DoF_{IAC}^{upper})^* = 2M$ when $K = 3$. An intuitive explanation on this result follows in the below. Let us first consider the IA case in the 3-user interference channel, where each of Tx and Rx has $M$ antennas. The two interferers can be considered virtually as a single one with the alignment operation, even while each receiver $k$ sees two interferers. For example, receiver 1 virtually regards transmitter 2 as the only interferer and receiver 2 virtually regards transmitter 3 as the only interferer, while receiver 3 virtually regards transmitter 1 as the only interferer. For a successful decoding, a sum of the simultaneous data streams for each combination in the above should be no more than $M$. Therefore, given an extreme assumption that interferences can be always aligned, the maximum total number of simultaneous data streams through IA in 3-user interference channel will be $3M/2$. Then, following the same principle, let us look into the IAC case for three interfering MACs. Also with IA operation, receiver 1 virtually regards the 2nd MAC as a massive interferer and hence, a sum of the simultaneous data streams for the 1st MAC and 2nd MAC should be no more than $M$. For receiver 2, since the interference effect caused by the 1st MAC has been subtracted by IC operation, it only regards the 3rd MAC as an interferer. Similarly, a sum of the simultaneous data streams for the 2nd MAC and 3rd MAC should be no more than $M$. For receiver 3, however, it will be different in the sense that the maximum of $M$ data streams can be transmitted, as no interferences exist due to the IC operation.

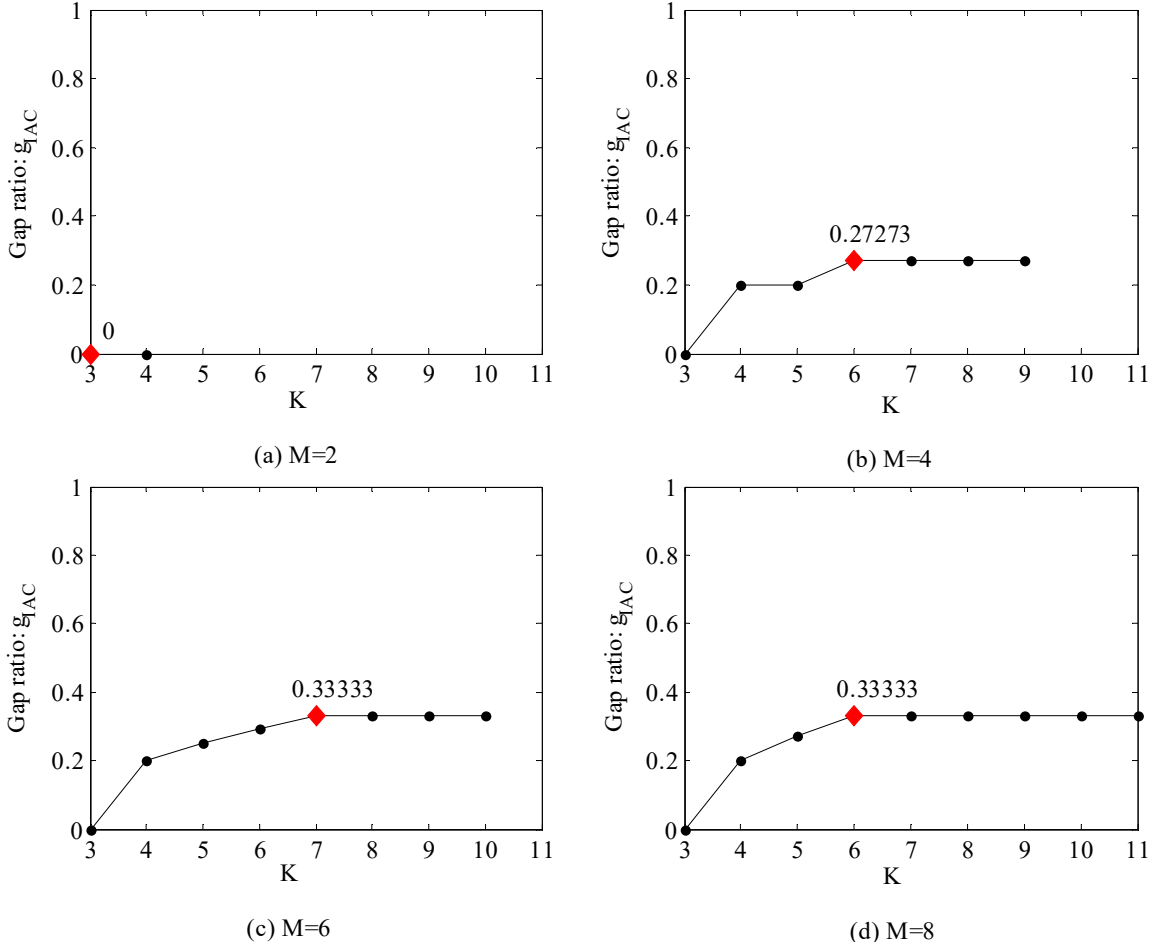

Fig 3. The gap ratio $g_{\text{IAC}}$ for $M = 2, 4, 6, 8$

Therefore, by combining the results of three MACs, an upper bound of $2M$ can be obtained. Finally, together with $DoF_{\text{IAC}}^{\text{CS}} = 2M$, it indicates that our proposed closed-form solutions can meet the upper bound when $K = 3$. This also illustrates that our proposed scheme has made a full use of the whole spatial dimension characterized by three interfering MACs.

Now, we present the numerical results to compare $DoF_{\text{IAC}}^{\text{CS}}$ to $\left(DoF_{\text{IAC}}^{\text{upper}}\right)^*$. To check their difference, we define the gap ratio as $g_{\text{IAC}} = \left[\left(DoF_{\text{IAC}}^{\text{upper}}\right)^* - DoF_{\text{IAC}}^{\text{CS}}\right] / \left(DoF_{\text{IAC}}^{\text{upper}}\right)^*$. As shown in Fig. 3, it is observed that $g_{\text{IAC}} = 0$ for $K = 3$, confirming that our proposed closed-form solutions have achieved the upperbound when $K = 3$. For all other cases, we observe that $DoF_{\text{IAC}}^{\text{CS}}$ can reach approximately 66.6% of $DoF_{\text{IAC}}^{\text{upper}}$.

Moreover, the curve $g_{\text{IAC}}$ increases yet eventually converges to a constant as $K$ increases, which indicates the maximum value of $g_{\text{IAC}}$ by a diamond mark in Fig. 3. We observe from these results that the capacity gap increases with both $K$ and $M$. However, our proposed closed-form solution can achieve at least 66.6% of the upper bound on the capacity, even for the large number of antennas. In fact, its performance can be guaranteed within 20% of the upper bound with a small number of K, e.g., $K \le 4$. Note that some values of $g_{\text{IAC}}$ are not shown, simply because IAC cannot be applied to some values of $K$.

## V. CONCLUSION

Following our earlier proposals on the closed-form IAC solutions for $(K, M, J)$ Gaussian interference MAC channel, this paper aims to derive the total DoFs that can be achieved by them. Instead of the specific configuration that has to be the same number of data streams for all transmitters, a general tuple of DoFs $\left\{d^{[j,k]}\middle|\forall j\in[1,N_k],\forall k\in[1,K]\right\}$ has been investigated. We have shown that regardless of $K$ and $J$, a total of $DoF_{\text{IAC}}^{\text{CS}}=2M$ can be achieved. The results also hold true for the $K$-user MIMO interference channel, which can be immediately reduced from the MAC channel. In the existing works on the total achievable DoFs obtained by IA in $K$-user MIMO interference channel [2,3], the maximum total DoFs of $2MK/(K+1)$ is achieved only by a specific configuration where each link has the same target DoFs. For a general tuple of DoFs, however, there only exists an upper bound on the total achievable DoFs, which given as ($2M-1$). In fact, this upper bound is not tight since it is obtained by easing the inequality constraint. As compared with these results, our closed-form IAC solutions improve the total achievable DoFs over IA for both specific and general cases. More interestingly, $DoF_{\text{IAC}}^{\text{CS}}=2M$ can be achieved only with $k_{\text{IAC}}=2$. It implies that our improvement can be achieved just by sharing the decoded signals of only two receivers through a backhaul link for cancellation. In other words, additional capacity has been obtained only by a reasonable amount of complexity.

We have evaluated our results by referring to the optimal IAC solutions. As it is NP-hard to derive the maximum achievable DoFs by optimal IAC solutions, its upper bound has been derived for comparison by relying on a necessary feasibility condition for IAC, which is just an extension of the existing IA result. As sufficiency of the necessary condition remains unknown, the tightness of the upper bound has not been proven until now. Furthermore, it is quite challenging to check the necessary condition for each subset of alignment equations. Therefore, we derive an upperbound by checking the complete set of alignment equations, which makes the upper bound generally looser than the one obtained by checking each subset. The numerical results have shown that the proposed closed-form IAC solutions can achieve the upper bound for the case with three interfering MACs, and can achieve approximately 66.6% of the upper bound for other cases with more than three MACs. As this rather large gap seems to be mainly attributed to the loose upper bound, our future work will be also focused on deriving a tighter upper bound.